\documentclass{llncs}

\usepackage[latin1]{inputenc}
\usepackage[T1]{fontenc}
\usepackage{amsfonts}

\def\jumpTo{{\small\sf back\,to}}

\def\et{,\ \ }
\newcommand{\reglecontrole}[4]{\makebox[2.5cm][r]{{\sf\small #1}}~$\displaystyle\frac{#2}{#3}~ #4$}

\def\oadymppac{OADymPPaC}

\begin{document}

\mainmatter
\title{On using Tracer Driver for External Dynamic Process Observation 
\thanks{This work has been partly supported by \oadymppac{}~\cite{oadimpac}, a French RNTL project.}
(extended abstract)}
%
%
\author{Pierre Deransart}
%
%
\institute{{\sc Inria} Rocquencourt, BP 105, 78153 Le Chesnay Cedex, France\\
\email{Pierre.Deransart@inria.fr}
}
\maketitle

\section*{Abstract}
One is interested here in the observation of dynamic processes starting from the traces which they leave or those that one makes them produce. It is considered here that it should be possible to make several observations simultaneously, using a large variety of independently developed analyzers.

For this purpose, we introduce the original notion of ``full trace'' to capture the idea that a process can be instrumented in such a way that it may broadcast all information which could ever be requested by any kind of observer. Each analyzer can then find in the full trace the data elements which it needs. This approach uses what has been called a "tracer driver" which completes the tracer and drives it to answer the requests of the analyzers.

A tracer driver allows to restrict the flow of information and makes this approach tractable. On the other side, the potential size of a full trace seems to make the idea of full trace unrealistic.

In this work we explore the consequences of this notion in term of potential efficiency, by analyzing the respective workloads between the (full) tracer and many different analyzers, all being likely run in true parallel environments.

To illustrate this study, we use the example of the observation of the resolution of constraints systems (proof-tree, search-tree and propagation) using sophisticated visualization tools, as developed in the project OADymPPaC (2001-2004).    

The processes considered here are computer programs, but we believe the approach can be extended to many other kinds of processes.

\section{Introduction}
One is interested here in the observation of dynamic processes starting from the traces which they leave or those that one makes them produce.  It is considered here that it should be possible to make several observations simultaneously, using a large variety of independently developed analyzers.

When one wants to observe a process, the practice is to instrument it for each type of observation which one wants to make on it. One thus implements a new "ad hoc" tracer for each analyzer, or one adapts and completes an existing one. Such a work can be largely avoided if one adopts from the start a general approach which consists to instrument it such that it produces what we call a "full trace". This unique trace will be useful then for all the later observations which one can plan to make. Each analyzer can then find in the full trace the data elements which it needs. This approach uses what was called a "tracer driver" which completes the tracer and drives it to answer the requests of the analyzers.

This approach is particularly tempting in practice as the full trace never needs to be completely expressed (the exchanges of information remain limited) and the work of the tracer implementation and of its driver is made once for all. The evaluation in terms of feasibility and performance remains however problematic.  

In fact, this approach allows to reduce the size of the trace emitted to a useful bare minimum and thus to speed up the whole process. In compensation, it allows to consider very large full traces. This has also a cost which grows with the size of the trace. Beyond a certain size, the production cost of the trace is likely to become prohibitory. It is precisely the question in which one is interested here. Until where can one go in this approach on a practical level without slowing down a process excessively and how? To get a more precise idea, one must take into account not only the time of trace production, but also how the trace will be used.

\vspace{3mm}
  In \cite{DLads05} and \cite{DLD05wlpe} the notion of tracer driver is presented and experimented is the context of finite domain constraint resolution \cite{LD04iclp}. But the question of the nature of the trace whose emission is controlled by the tracer driver is not directly tackled. In this work we explore the consequences of this notion in term of potential efficiency, by analyzing the respective workloads between the (full) tracer and many different analyzers, all being likely run in a true parallel environment. 

In this work, we introduce the notion of ``full trace'' to capture the idea that a process can be instrumented in such a way that it may broadcast all information which could ever be requested by any kind of observer. We then analyze the nature of the work of the tracer and the driver, and the distribution of the functions between the tracer and its driver on the one hand, and the analyzers on the other hand. This allows us to better estimate how powerful, useful and efficient the concept of full trace can be, provided it is accompanied by the right architecture of all involved components.

To illustrate this study, one will take the example of the observation of the resolution of constraints systems (proof-tree, search-tree and propagation) using sophisticated tools for visualization, according to the method developed in the projects DiSCiPl (1997-2000) \cite{DISCIPL:LNCS} then OADymPPaC (2001-2004) \cite{oadimpac}. This field is of particular interest because the traces include the representations of complicated and potentially bulky objects, and the computations (evolution of the domain of the variables) are at the same time logical and stochastic. The constraints systems, because of the complexity of their resolution, are very close to true complex systems.

\vspace{3mm}
In this extended abstract we present successively the concepts of full trace and its incremental ``compressed'' version, then the question of their semantics. We analyze finally the problem of the distribution of work between a driven tracer and external analyzers which work only with the useful trace flow which is provided to them by their requests.

\section{Full Trace}
We  introduce here the concept of  {\bf full trace}. By definition the full trace of a process contains all one may like to know about it during its execution (this includes likely a description of the process itself) . A process is characterized at a given moment $t$ by a state $S_t$. It does not enter in the framework of this article to define what is exactly such a state. It will be supposed only that it can be described by a finite set of parameters $p_n$ for the $n_{th}$  parameter, and $p_{n,t}$ for its value at moment $t$. The concept of "moment" will be specified hereafter. A current state will be denoted by the list of the values of its parameters. It is also assumed that the transformation of a state into another is made by "steps" characterized by an action $a$. The set of actions performed at moment $t$ is labelled $a_t$.

So defined, the concept of full trace seems not to have any application. In practice, there are only approximations; but the important thing here, is to admit that, whatever is the level of details with which one wishes to observe a process, there is always a threshold which makes it possible to define such a trace. One can consider that in the case of a program, it acts of a more or less thorough instrumentation which produces a trace, in other words a program augmented with a tracer.

\begin{definition} [Virtual Full Trace].  
An virtual full trace is an unbounded sequence of trace events of  the form {\bf $e_t: (t, a_t, S_{t+1})$} comprising the following elements:
\begin{itemize}
  \item $e_t$: unique {\bf identifier} of the event.  
  \item $t$: {\bf chrono}. Time of the trace. It only varies by unit values and is always increasing. To distinguish from the time of the observed processes or of the analyzers which may not be monotonous compared to the chrono.  
  \item $S_t = p_{1, t}..., p_{n, t}$: {\bf parameters} at chrono $t$. In a trace event  the parameters are called  {\bf  attributes} and $S_t$ the  {\bf  full current state}. The parameters may describe objects or actions performed to reach the new state. $S_t$ is the last ``observed'' state before the event $e_t$ occurs. The parameters (or attributes) correspond to the new reached state $S_{t+1}$.
  \item $a_t$: {\bf action} an identifier of the set of actions characterizing the step from the state $S_t$ to the state $S_{t+1}$.
\end{itemize}
\end{definition}

Any trace effectively produced by a process can be regarded as a partial full trace. In practice, one "sees" only partial traces which start after the moment 0 where the process observed is in a presumably initial state $S_0$. We will limit ourselves in this summary to only a single example of trace: the trace of the proof-trees in Prolog systems based on standard Prolog \cite{gnuprolog}. This trace is far from being a full trace (many useful information, even easily available at any moment are not represented there). Here is an extract:

\begin{verbatim}
      1    1  Call: '$call$'(bench(2))
      2    2  Call: bench(2)
      3    3  Call: 2>0
      3    3  Exit: 2>0
      4    3  Call: _182 is 2-1
      4    3  Exit: 1 is 2-1
      5    3  Call: bench(1)
      ....
\end{verbatim}

This is a Byrd's trace  \cite{byrd80lpw}, adopted in the majority of the Prolog systems. The moment "0" corresponds to the launching of the resolution of a goal. Each trace event corresponds to a stage in its resolution. An event contains the following information: two attributes which are indication of depths (the first, the depth in the search-tree and the second in the proof-tree), a  "port" which corresponds to the action which made it possible to reach this stage. Thus the port {\bf call} corresponds to a call to the indicated subgoal and the installation of the subgoals to solve, and {\bf exit} to the success of   the subgoal. There are other ports (a total of 5 in the case of GNU-Prolog) not detailed here. The first "call" defines the launching of the first goals (if there is only one of them). If the resolution terminates, the trace is finite, unbounded otherwise. The last attribute gives the  subgoal to solve. This trace does not comprise identifiers nor chrono; the chrono corresponds to the sequential order of emission of the trace events and the resulting chrono plays the role of an identifier.

It is interesting to note that the objective of such a trace is to display all the steps of evolution of a proof-tree until obtaining all the possible proofs (the port {\bf fail} corresponds to a failure and {\bf  redo} to  a nondeterministic goal where a new resolution will be tried), and to some extend to describe also the search-tree. However, the (partial) proof-tree is never explicitly described (not more than the search-tree). This trace thus does not provide the parameters of interest directly (the partial proof-tree). This brings us to the following observation. The parameters of the full state are not given explicitly in this trace, but only some attributes (port and goal) which possibly make it possible to find it. We will reconsider this point later. These attributes thus give only an incremental information which makes it possible to obtain the new proof-tree after a resolution step. We will say that the trace is "incremental". This leads us to define the following particular traces which we define here without more details.

\begin{definition} [Discontinuous, Effective, Incremental Full Trace].  

 \indent A trace is {\bf discontinuous} if it is a (full) trace which contains successive events whose variation of the chrono may be higher than a unit. There are "holes" in it, either because the emission is discontinuous, or because the observing process "listens" only occasionally to it. Notice that the discontinuity concerns only the trace emission and not its extraction; in other words moment 0 corresponds to the activation of the tracer. 

 \indent A trace is {\bf effective} if it is a (full) trace of the form {\bf $e_t: (t, A_t)$} and such that, starting from the knowledge of $(S_t, A_t)$ one can deduce $(a_t, S_ {t+1})$. $A_t$ denotes a set of attributes. The effective trace is the trace emitted by a tracer, that which is actually ``visible''. The virtual full trace is a particular case of effective trace where the attributes $A_t$ are the action label $a_t$ and the parameters $S_t$. Another particular case is the full ``incremental'' trace. 
 
\indent
A trace is {\bf  incremental} if  the attributes are such that only the changes affecting the current state are noted. It has the form {\bf $e_t: (t, Delta_t+)$} where  $Delta_t+$ contains the description of the actions which modify the values of the parameters of the moment $t$. To remain a full trace, this trace must satisfy the following condition: starting from the knowledge of $(S_t, Delta_t+)$ one can deduce $(a_t, S_ {t+1})$.
\end{definition}

In this extended summary, when the distinction is not absolutely necessary, one will not distinguish between virtual and effective traces, and one will speak indifferently about parameters or attributes. 

Practically all the traces are incremental (thus it uses attributes), as the preceding example illustrates it well, because the emission of a full state in each trace event would be obviously prohibitive. In fact the size which would take the events would be much too high and the trace would be extremely redundant. The condition imposes simply that one can retrieve the full trace starting from the transmitted attributes and from the preceding full state. In practice the observed processes will produce partial traces only. In this case the retrieval of the full state is impossible. If one wants to find a full trace, or at least a more complete one, it will be necessary "to ask" the tracer to provide at least one full state. If the  observing process needs to take into account only a partial state, that can be sufficient to enable it to maintain a consistent partial state. On the other hand if it needs, at a given moment, to know a full state, or at least a more complete state, it will have ``to ask'' the tracer to provide him at least a full current state or a part of it. 

Practically all the traces are discontinuous, even if often each part of them (necessarily finite) can be regarded as a finite single trace.
This also raises the problem of the knowledge of the initial state $S_0$ in which the observed process was at the moment of the initial trace event (chrono equals to 0) and thus of the communication to the analyzers of the full initial state $S_0$, before any event of trace. 

These two reasons justify that one is interested in the manner of obtaining such a state. The current value of a parameter may exist just as it is in the process and requires only a small computation to extract it. On the other hand it may not exist and should require a partial re-execution  of the process (this capacity is used in the analyzers of the CHIP environment of Cosytec \cite{DISCIPL:LNCS} or CLPGUI \cite{oadimpac}). This obliges to stop the execution of the observed process (at least when that is possible), and to give the ability to the observing process to stop and resume  the observed process. This leads to the idea sketched in the introduction of synchronization primitives between processes, a kind "image freezing" which makes it possible to complete the information at any moment, according to the need. This also results in the need for a function "access to the current state".

It is important to note that with an effective or incremental trace, keeping a full current state is then in charge of the observing process and not of the observed process. This assumes however that the observed process contains at least recovery points in which full current state will be maintained  and accessible, so as to allow the observing  process to resume the trace and to be able to restart from a full current state.
It is important to note that with a full effective trace, different from the full virtual trace, to keep a full current state in the observing process will be an additional charge of this process itself, because the tracer of the observed process then does not have any more obligation to calculate the requested parameters explicitly (only the requested attributes are computed). This supposes however that the  observed process contains at least ``recovery points'' in which a full current state will be preserved and accessible. This may allow the observing process to resume the trace and to be able to restart from a full current state. This aspect is not treated here. 

\section{Observational Semantics}
One is interested in this section in the semantics of the full trace, called here "observational semantics".

The trace does not explain anything.  It is only a collection of facts. The question arises however to understand  the trace. 
For that one needs two levels of semantics. The first level corresponds to the description of the actions and objects appearing in the parameters and attributes of the traces, i.e. a semantics of the observed data. The second level corresponds to a kind of trace ``explanation'', i.e. a semantics describing how the values of the parameters at the moment $t$ are derived from the values of the parameters of the moment $t-1$ and how the actions $a_t$ are selected. 
Clearly this means that one has a model of the process which accurately describes the evolution of the trace between two events. The form of this semantics is the subject of another work in progress.

``To read'' a trace, one needs only the first level of semantics. 
Only the relations between the parameters in the same state or attributes in the emitted trace events must be known. In the example of the Prolog trace, the properties of the object "proof-tree" must be known to understand the relation between the depth of the tree and the tree itself. The semantics of the trace (full and/or incremental) thus contains the description of the relations between the parameters which relates the semantics and the produced attributes. But to understand the trace comprehensively, it is necessary to go further in the knowledge of the process itself.

What we call here "observational semantics" (OS) is the semantics of the tracer 
(the first semantics, although being part of the OS, can be seen like a ``semantics of the trace''). It uses the semantics of the objects and actions of the full trace and can be seen as an abstract semantics in the sense of Cousot \cite{cousot}. 
It is a kind of natural semantics \cite{kahn87} which can be expressed by a finite set of rules of the form 

$a:\ Condition(S) \ \rightarrow \  S'\ =\ a(S)$ 

such that  any event of trace $e_t: (t, a_t, S_{t+1})$ can be obtained starting from the state $S_t$ and the action $a_t$ by application of the rule $a$ of  the OS. The (set of) actions $a$ performed at chrono $t$ is denoted $a_t$. This semantics can take the form of a structural operational semantics \cite{plotkin81} or of an evolving algebra \cite{gurevitch91} and being more or less refined, "big-step" or "small-step" \cite{Lucas99}. It is in fact very tempting to have a complete semantics for the tracer in order to allow a clear implementation of it.

Coming back to the example of Prolog, a tracer augmented with the trace of constraints resolution (proof-tree, search-tree, labeling-tree and propagation), called Codeine, has been implemented in this manner on GNU-Prolog \cite{langevine-iclp03}. An abstract model has been defined \cite{LDD04lnai}, which has been  then implemented on several solvers \cite{oadimpac}. This made it possible to meet two principal objectives: portability of the analyzers whose input data are based on the full trace, and robustness of the tracers whose implementation is guided and improved by a good methodological approach \cite{DLD03aadebug} based on a pre-specified rigorous semantics.

It is useful however to note that a complete semantics of the trace, which would be a complete formalization of the observed  process, is almost impossible in practice, because of the degree of refinement that would imply. For example, an attribute included in many traces is the CPU time consumed by the process since the "beginning" of the trace emission. To formalize its variations, in fact issued from the host system, would amount introducing into the observational semantics a model of the system in which the process is executed.

Finally an OS should not be confused with a complete operational semantics which would make it possible to recall the course of the process starting from the sole initial state and its rules. The current full state is not sufficient to know which rule may apply. 
For example, the model developed for finite domain constraints resolution described in \cite{langevine-iclp03} contains a set of operational rules such as at least one rule can be applied at each state, but whose conditions are not always precise enough to decide how it can be applied. So is the \jumpTo{} rule, depicted in the Figure~\ref{cprule}. 

\begin{figure}[t]\small
\noindent
\def\et{,\ \ }
\reglecontrole{\jumpTo{}}{\phantom{xx} l \in \mathrm{dom}(\Sigma) \;
  \wedge \;choice\hbox{-}point(\Sigma(l))\phantom{xx} }{{\cal N} \gets l, {\mathbb S} \gets
  \Sigma(l)}{}
\caption{A rule of the abstract model of Prolog Proof-Tree displayed in \cite{langevine-iclp03}}
\label{cprule}
\end{figure}

It contains a condition $\phantom{xx} l \in \mathrm{dom}(\Sigma) $ which must be satisfied, meaning that the new current node belongs to the search-tree, but nothing says which node must be selected. The sole knowledge of the full current state (which includes the current search-tree) is not sufficient. The knowledge of the current trace event is necessary to know which is this new node and thus to know how this rule  is in fact instantiated. 

\section{Tracer Driver and Evaluation}
The idea to use a tracer driver was proposed by M. Ducassé and J. Noyé since the origin of Prolog \cite{ND94} in the context of logic programming, then regularly developed and tested in various environments \cite{dag-mireille03}. The originality of this approach consists in proposing that the full trace should not be communicated in its integrity to an analyzer, but that this one receives only the part of the trace which relates to it. Instead of  broadcasting "urbi et orbi" a full trace that any analyzer would have the task to filter, filtering is performed at the source level, i.e. by the observed process which behaves then like a server of traces. The analyzer is a client which  restricts himself to indicate to the observed process the trace which it needs indeed. The complete separation of the observed and observing processes then brings to consider an architecture server/client with exchanges of information: the client indicates to the server the trace it needs and the server provides him only what it requires. The filtering of the trace is no longer carried out by the analyzer, but by the observed process. It is the task of the tracer driver to perform all requested filtering and to dispatch the traces requested by the clients. This architecture as well as the exchanges between the processes have been described in particular in \cite{oadimpac} and \cite{DLads05}. We will not go more into the details in this extended abstract. In addition, only the aspects related to the possibilities of modulating the traces emitted according to the needs of the analyzers will be considered; the possibility previously mentioned to synchronize the processes will not be considered here.

This approach allows to reduce the size of the trace emitted to a useful bare minimum and thus to speed up the whole process. In compensation, it allows to consider very large full traces. This has also a cost which grows with the size of the trace. Beyond a certain size, the production cost of the trace is likely to become prohibitory. It is precisely the question in which one is interested. Until where can one go into this approach on a practical level without slowing down a process excessively and how? To get a more precise idea, one must take into account not only the time of trace production, but also how the trace will be used.

\vspace{3mm}
One could think at first sight that the fact, for the observed process, to have to produce a full trace makes this approach unrealistic. The server is indeed slowed down by the simple fact of having to compute a great number of parameters which will perhaps never be used. This approach would be thus very penalizing for the process instrumented with an expensive tracer whenever only a weak portion of the full trace would be used.
On another side, and precisely in this case, a considerable economy is realized because of filtering at the source, and the transmission of a limited trace whose costs of coding and diffusion are then extremely reduced. The idea is that in practice the fact of emitting only one small part of the trace, filtered at the source, compensates for the over cost of work mainly related to the updates of all the parameters of the full trace in the analyzed process.

On the other hand, if many analyzers are activated simultaneously and use at a given moment a trace equivalent to a full trace, the problem to produce or not the full trace does not arise any more (it must in any case be produced), and the question about the interest of using a tracer driver is worth to be posed. We show that even in this case the driver can save time. 

Additionally another type of economy must be considered which is complementary to the previous one: the use of an incremental trace as an effective trace instead of the virtual full trace. In this case there is no loss of information (the equivalent of the full trace is still transmitted), but there is a reduction of the amount of data transmitted (a kind of ``data compression'' is used in order to limit the size of the emitted data flow). In this later case the tracer performs a kind of ``compression'' and the analyzer a kind of ``decompression''  and both task must be taken into account in the analysis of the workloads. It will be shown that the performances of the tracer and the analyzers can be influenced more by this kind of load rather than by the size of the full trace. 

\vspace{3mm}
In order to be more precise, it is necessary to analyze the repartition of work within the various processes.  

The times to take into account for a detailed performances analysis are times concerning the process itself, its tracer and its pilot on the one hand, and times concerning one analyzer on the other hand (one must suppose here that, if there are several analyzers, they are running with true parallelism): 

\noindent
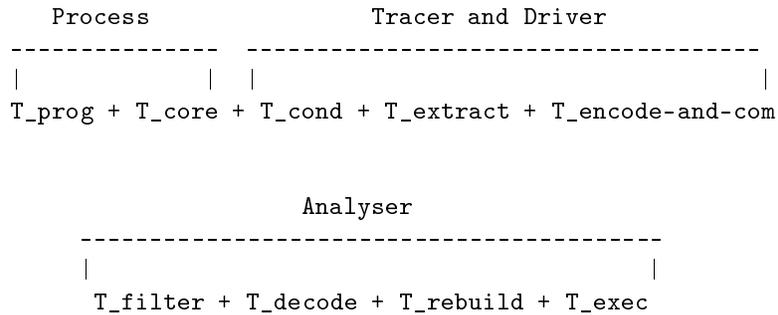
\begin{verbatim}
        Process                Tracer and Driver                
     ---------------  ------------------------------------- 
     |             |  |                                    |
     T_prog + T_core + T_cond + T_extract + T_encode-and-com 

 
                          Analyser                
          ------------------------------------------ 
          |                                        |
           T_filter + T_decode + T_rebuild + T_exec
\end{verbatim}

From the side of the process, tracer and driver: 
\begin{itemize}
\item $T_{prog}$: time devoted to the execution of the not instrumented program (or instrumented but with deactivated tracer).
 \item $T_{core}$: additional execution time of the process once instrumented to produce the full trace and with activated tracer. It is a time devoted to the construction of the elements necessary to a likely later extraction of the parameters of the current state. This time is largely influenced by the size of the full trace. 
At this stage all computations must be performed because, in the presented approach, one considers that it must be possible to produce the full current state at any moment (in case in particular of discontinuous trace). This time is related to the form of the full trace only and does not depend on what will be emitted. If all the parameters of the full trace are already part of the process, this time is just null. 
 \item $T_{cond}$: time of checking of the conditions defining the traces to be emitted for each analyzer (filtering). This time is null if there is no filtering (emission of the full trace). 
 \item $T_{extract}$: computing time of the parameters requested during or after filtering. 
 \item $T_{encode-and-com}$: time  for formatting the trace (encoding), possible compression and emission.
\end{itemize}
Time specific to the driver corresponds to $T_{cond}$. Other times, namely, $T_{core}$, $T_{extract}$ and $T_{encode-and-com}$ can be regarded as times related to the tracer.

It should be noticed here that, in this approach, the driver acts only on the choice of the parameters and the attributes to be contained in the effective trace, i.e. on the communicated information. It does not have the possibility of influencing the form of the attributes, for example the degree of ``compression'' of the information. In fact, this ``compression'' is coded here in the form of the attributes. The nature of the attributes (incremental information or not) is part of the tracer and it cannot be modified or adapted through the tracer driver. On the other side any additional compression algorithm used to reduce the size of the information flow belongs to the stage of ``encoding''. Generalizing this idea, one could also consider the possibility to put into the trace more abstract attributes, adapted to some more specific use, such that the whole trace has a reduced size. The corresponding attribute computation time would thus be related to the ``extraction'' stage. This generalization is not studied here. 

\vspace{3mm}
From the side of the analyzer: 
\begin{itemize}
 \item $T_{filter}$: time of filtering by an analyzer. This time is null if filtering is performed at the source (the trace events sent to a particular analyzer can indeed be tagged at the source). In fact by precaution but mainly because of lack of implemented tracer driver, many external analyzers will filter again the trace. However it is not necessary to consider this case here. 
\item $T_{decode}$: time of decoding of the received trace. This time is impossible to circumvent, as the time of coding and communication $T_{encode-and-com}$ as well. It must be compared with the $encode$ part of $T_{encode-and-com}$. It cannot be completely eliminated. If compression/decompression algorithms are used, it is because it is considered that, even if both times are cumulated, one will save a substantial amount of time over the transmission. 
\item $T_{rebuild}$: time of rebuilding the full trace starting from the effective trace (computation of the current parameters starting from the emitted attributes). This time must be compared with $T_ {extract}$, the computation time of the attributes. It is considered that such a time, even if cumulated with $T_ {extract}$, equilibrates most favorably with the cost of trace emission.
\item $T_{exec}$: execution time of the proper functions of the analyzer. With very sophisticated analyzers (as those used for data analysis for example), this time can become so important that it makes negligible the one corresponding to the trace production. 
\end{itemize}

Example: Codeine, described in \cite{langevine-iclp03}, implements the generation of a full trace for the analysis of constraints resolution, as an extension of GNU-Prolog. The current state $S_t$ contains (among other attributes):  proof-tree, search-tree, constraints state and variables with their domains. For a complete description of the full trace see \cite{oadimpac} (called in this project "generic trace"). Even if Codeine does not generate a full trace (as full as it could be possible), the Codeine trace is considerably richer than the simple Byrd's trace of  GNU-Prolog which is strictly contained. Only an incremental trace is generated, and only the proof-tree and the constraints and variables states can be rebuilt later starting just from the current state of the process (the cost of extraction is reduced to the cost of the proper data management realized by the tracer). To obtain the current search-tree the process would have to be re-executed partially, therefore it would be necessary to freeze its current execution. The cost of management of a permanently accessible search-tree at any moment would be clearly intractable because of its size. On the other hand an analyzer can, using the produced incremental trace,  maintain all these objects permanently (obtaining $S_t$ or those parameters useful for it).   Codeine also contains a tracer driver such that the definitions of the traces to be emitted (specification of the emitted trace) are stored in a data file which must be provided before the process starts.

\vspace{3mm}
Times are distributed as follows:

From the side of the process, tracer and driver: 
\begin{itemize}
  \item   $T_{prog}$: execution time of GNU-Prolog with "switches" of the tracer (a small part of $T_{core}$ of negligible duration in general). 
 \item $T_{core}$: time of construction of the parameters of the full Codeine trace (collecting of all data useful to extract the full trace); a kind of  GNU-Prolog  plug-in. 
 \item $T_{cond}$: time of filtering the full trace to select all the requested traces.  
\item $T_{extract}$: computing time of the attributes corresponding to the parameters requested during filtering. 
\item $T_{encode-and-com}$: computing time of the attributes corresponding to the requested parameters, encoding  (XML format or Prolog term) of the emitted trace, and emission time of the incremental trace. 
 \end{itemize}

From the side of the analyzers:  experiments in the framework of the OADymPPaC project with sophisticated analyzers (intensive visualization of graphs, visual data analyses) revealed the following costs. 

\begin{itemize}
\item $T_{filter}$ and $T_{decode}$: both times are intricate in a syntactic analysis module (XML) of the full trace. Re-filtering, a part of which could have been avoided. But the ``driven tracer'' approach  could not be taken into account in the analyzers built during this project. 
\item $T_{rebuild}$: time of construction of the parameters of the full trace (variables domains, active constraints set, search-tree ...). This time grows (non estimated factor) according to the size of the data, sometimes related to the size of the trace, with a considerable slow down of the analyzer during the ``reading'' of the trace . The low speed of an analyzer may cause, in the case of analysis of the trace ``on the fly'', a strong slow down of the observed process. 
 \item $T_{exec}$: time of construction of the objects to be visualized (graphs, data tables). These times can grow exponentially according to the size of the data. The efficiency of the used algorithms is crucial here. This can also cause a slow down of the process, and pleads in favor of a preliminary treatment of information before transmission (for example, selection of distinguished nodes to put in the trace to reduce the size of a drawn graphs, or to collect groups of variables as a unique attribute to reduce the number of lines in a matrix). 
 \end{itemize}

It was been shown experimentally  in \cite{DLD05wlpe} that the behavior of the Codeine tracer with the full trace above compares favorably with the behavior of GNU-Prolog with the Byrd's trace only. Furthermore, the filtering realized by the tracer driver does not prejudices  the performances.

We give here a theoretical justification to this result, showing that this approach, already justified experimentally, can also be justified theoretically.

In \cite{DLads05} L. Langevine observes that filtering the traces all together is more effective than to filter them one after the other. This comes from the fact that running several automata together may be more efficient than running one only. Indeed the essence of filtering relates to simplified conditions whose role is to select first  the trace events containing  the ports requested by an analyzer. A finer additional filtering will be thus carried out thereafter but on a number of events much more reduced. One can admit that this first filtering relates to a trace whose language corresponds to a regular language. It is then possible to consider that each filter is itself a regular expression whose recognition on the full trace can be done using a non deterministic finite-state automaton. Filtering corresponds then to the recognition task by a union of as many finite-state automata than there are active analyzers requesting a trace. However the resulting automaton, once optimized, can be much more efficient than the most efficient of the automata associated with a single analyzer (the union of the automata can be more efficient in terms of computation steps than only one of them \cite{HU79}). As these operations of filtering are extremely frequent, because they apply to all the trace events, the speed up can be considerable.

\vspace{3mm}
We are now in position to analyze the respective workloads.

First observe that respective times of both sides may be considered as cumulative or not, depending whether the respective process are run sequentially or with true parallelism. In the later case, if all process are run on different processors (a situation which can be considered with such an approach), only the slowest process must be taken into account to evaluate the execution time of the whole system.

\begin{itemize}
\item $T_ {prog}$ and $T_ {core}$ on one side, $T_ {exec}$ on the other side. 

These times correspond to times specific to the tracer and the analyzers (the slowest analyzer has the main influence on the execution time). These times are incompressible and $T_ {core}$ depends only on the size of the virtual trace.

\item $T_ {cond}$ on one side and $T_ {filter}$ on the other side.

At least one of these times must be null, and, if the filtering is at the source, the performance can be improved. In any cases this time appears negligible, whatever is the size of the full trace, compared with the other times. 

\item $T_ {extract}$ on one side and $T_ {rebuild}$ on the other side.

These times correspond respectively to the computation of the attributes from the parameters  and reciprocally. If one can reduce the extraction time thanks to the  filtering (considering the low number of selected trace events and attributes, and the fact that the most of the parameters computation work is already included in  $T_ {core}$), the time of ``rebuilding'' can be very important and will probably be significantly greater than the extraction time.

\item $T_ {encode_and_com}$ on one side and $T_ {decode}$ on the other side.

These times can be important, but the interest lies in the fact that the profits realized on the communication time (reduced emitted volume) largely compensates the coding/decoding times. 
\end{itemize}

To summarize, the tracer and the slowest analyzer are the main factors influencing the whole performance and they can increase the workload of both sides considerably. However the use of a tracer driver and of techniques of ``trace compression''  make it possible to compensate partially, but sometimes in very effective manner, the over-costs related to the use of a full trace.

\section{Conclusion}

We introduced the concept of full trace in order to take into account  the multiple possibilities of analysis of a dynamic process. The analyzed process is instrumented with, in addition to its tracer, a  trace driver, and the analyzers have the possibility of addressing orders to the driver. An architecture client/server is then considered to describe the interactions between the processes analyzers and the analyzed process. This enabled us to tackle the problem of the evaluation of this approach in terms of efficiency.

We tried to appreciate how the introduction of a tracer driver could improve the global efficiency of the analysis of a dynamic process using several analyzers observing the process in a simultaneous way. We initially observed that a full trace could be particularly expensive to produce, but that part of this cost could be transferred, without loss of capacity of analysis (the full trace can be retrieved by the analyzers), on the analyzers. The most unfavorable case in term of efficiency corresponds to the situation where the equivalent of a full trace must be built, extracted and emitted. In this case, the full trace is in fact a union of all the reduced traces requested by several analyzers. We showed that, even in this case, a filtering realized by a tracer driver was able to bring an important benefice.
Our observations, during the projects DiSCiPl \cite{DISCIPL:LNCS} and OADymPPaC \cite{oadimpac}, using very sophisticated analyzers as  powerful tools for visualizations as well  (\cite{baudel04visCP} \cite{gjf04flairs}), also showed that the limits of performance came often more of the analyzers that tracer, even with a full trace.

\vspace{4mm}
This study opens finally on a series of questions:

\begin{itemize}
  \item   
{\bf  How deep is it possible to implement a very broad full trace?  } Of course the limits of the approach are obvious: the concept of full trace is meaningful only with regards to a family of possible analyses, well known and selected in advance. Nothing guarantees a priori that an additional instrumentation of the observed process will never be necessary. But beyond this aspect, the interesting question relates to the feasibility of the implementation of a very fine grained full trace. On one side indeed, one will be able to compensate for certain production costs of such large trace, by emitting  an effective trace of limited size; but on the other side, the analyzers (whose use can be temporary or exceptional) will be loaded with most of time of development/re-building of the full trace (necessary, even if it uses only a part of it) and will occasionally slow down the observed process.

\item   
{\bf   Which interaction language to use and which dialogue between the driver and the analyzers?  }
This question has been little tackled here and mainly remains out of the scope of this article. The tendency however is with the use of a language like XML. It is the way chosen by the OADymPPaC project, and the possibility of reducing the flow of trace to its bare minimum encourages this. Nevertheless our experiments showed that the communication needed to be optimized by combining  usual methods of data compression with specific trace compression like using ``incremental attributes'',  but also by introducing more abstract attributes in the trace.
This leads to the idea that the dialogue between the involved processes, limited here to the choice of the trace events and the attributes in the trace, and capacities of synchronization, must be extended such that it allows also to influence the design of the attributes themselves.

\item 
Finally a crucial question related to the comprehension of the trace. {\bf  How to understand a trace, or how to describe its semantics ?   }
If the semantics of the trace (and of the tracer), or at least a large part of it, is given a priori (because one has a model for the observed process), then the comprehension of the trace as well as the implementation of the tracer are largely facilitated. In the opposite case - and it is the case for many natural or artificial complex processes - one has only vast traces to study. Even in the fields of  programming languages where semantics seems better controlled a priori, one tries to analyze a program behavior by trying to understand  its traces. Thus one sees stinging the usefulness of general techniques based on data mining \cite{DDG05} or on Web mining \cite{csmr2005} for such purposes. However one must recognize that any full trace will probably always include portions escaping any kind of description based on formal semantics.
\end{itemize}

\bibliographystyle{splncs}

\begin{thebibliography}{10}

\bibitem{oadimpac}
Deransart, P., {\& al}:
\newblock {Outils d'Analyse Dynamique Pour la Programmation par Contraintes
  ({OADymPPaC})}.
\newblock Technical report, {Inria Rocquencourt} and {\'Ecole des Mines de
  Nantes} and {Insa de Rennes} and {Université d'Orléans} and Cosytec and ILOG
  (2004) Projet RNTL. {\tt http://contraintes.inria.fr/OADymPPaC}.

\bibitem{DLads05}
Langevine, L., Ducass{\'e}, M.:
\newblock {A Tracer Driver for Hybrid Execution Analyses}.
\newblock In Press, A., ed.: Proceedings of the 6th Automated Debugging
  Symposium. (2005)

\bibitem{DLD05wlpe}
Langevine, L., Ducass{\'e}, M.:
\newblock {A Tracer Driver for Versatile Dynamic Analyses of Constraint Logic
  Programs}.
\newblock In Serebrenik, A., Muñoz-Hernandez, S., eds.: Proceedings of the 15th
  Workshop on Logic-based Methods in Programming Environments ({WLPE'05}), a
  pre-conference workshop of {ICLP}'05, Sitges, Spain (2005) Computer Research
  Repository cs.SE/0508105.

\bibitem{LD04iclp}
Langevine, L., Ducass{\'e}, M.:
\newblock {A Tracer Driver for CLP: Debugging, Monitoring and Visualization of
  an Execution from a Single Tracer}.
\newblock In Demoen, B., Lifschitz, V., eds.: Proc. of the 20th Int. Conf. on
  Logic Programming, (ICLP'04). Number 3132 in LNCS, Saint-Malo, France (2004)
  462--463

\bibitem{DISCIPL:LNCS}
Deransart, P., Hermenegildo, M., Ma{\l}uszy\'nski, J., eds.:
\newblock Analysis and {V}isualisation {T}ools for {C}onstraint {P}rogramming.
\newblock Number 1870 in LNCS. Springer Verlag (2000)

\bibitem{gnuprolog}
GNU-Prolog:
\newblock {A CLP(FD) System Based on {Standard Prolog (ISO)} developed by {D.
  Diaz}} (2003) http://gprolog.sourceforge.net/ Distributed under the GNU
  license.

\bibitem{byrd80lpw}
Byrd, L.:
\newblock Understanding the control flow of {P}rolog programs.
\newblock Logic {P}rogramming {W}orkshop (1980)

\bibitem{cousot}
Cousot, P.:
\newblock {Abstract Interpretation}.
\newblock {Technique et Science Informatiques} \textbf{19} (1994)  1--9

\bibitem{kahn87}
Kahn, G.:
\newblock {Natural Semantics}.
\newblock In Brandenburg, F.J., Vidal-Naquet, G., Wirsing, M., eds.:
  Proceedings of STACS, Springer (1987) also INRIA, RR 416, Natural Semantics
  on the Computer.

\bibitem{plotkin81}
Plotkin, G.:
\newblock {A Structural Approach to Operational Semantics}.
\newblock TR DAIMI FN 19, Computer Science Department, Aarhus University,
  Aarhus, Denmark (1981)

\bibitem{gurevitch91}
Gurevitch, Y.:
\newblock Evolving algebras, a tutorial introduction.
\newblock Bulletin of the European Association for Theoretical Computer Science
  \textbf{43} (1991)  264--284

\bibitem{Lucas99}
Lucas, S.:
\newblock {Observational Semantics and Dynamic Analysis of Computational
  Process}.
\newblock RR 00/02, {Ecole Polytechnique, Laboratoire d'Informatique, LIX},
  Palaiseau (France) (2000)

\bibitem{langevine-iclp03}
Langevine, L., Ducass\'e, M., Deransart, P.:
\newblock A {P}ropagation {T}racer for {GNU}-{P}rolog: from {F}ormal
  {D}efinition to {E}fficient {I}mplementation.
\newblock In Palamidessi, C., ed.: Proceedings of the 19th International
  Conference on Logic Programming, ICLP'03, Mumbai, India (2003)

\bibitem{LDD04lnai}
Langevine, L., Deransart, P., Ducass\'e, M.:
\newblock {A Generic Trace Schema for the Portability of {CP(FD)} Debugging
  Tools}.
\newblock In Apt, K., Fages, F., Rossi, F., P., S., Vancza, J., eds.: Recent
  Advances in Constraints. Number 3010 in Lecture Notes in Artificial
  Intelligence.
\newblock Springer Verlag (2004) Selected papers of the Joint ERCIM/CoLogNET
  International Workshop on Constraint Solving and Constraint Logic
  Programming.

\bibitem{DLD03aadebug}
Ducass{\'e}, M., Langevine, L., Deransart, P.:
\newblock Rigorous design of tracers: an experiment for constraint logic
  programming.
\newblock In Ronsse, M., Bosschere, K.D., eds.: Proceedings of the 5th
  International Workshop on Automated and Algorithmic Debugging, AADEBUG'03,
  Ghent, Belgium (2003) Computer Research Repository cs.SE/0310042.

\bibitem{ND94}
Ducass\'e, M., Noy\'e, J.:
\newblock {Logic Programming Environments: Dynamic Program Analysis and
  Debugging}.
\newblock The Journal of Logic Programming \textbf{19/20} (1994)  351--384

\bibitem{dag-mireille03}
Ducass{\'e}, M.:
\newblock {Trace Schemata for (Multi-language) Dynamic Analysis}.
\newblock In Choi, J.D., Ryder, B., Zeller, A., eds.: Dagstuhl Seminar 03491,
  "Understanding Program Dynamics", Saint-Malo, France, Dagstuhl, Germany
  (2003)

\bibitem{HU79}
Hopcroft, J.E., Ullman, J.D., eds.:
\newblock {Introduction to Automata Theory, Languages, and Computation}.
\newblock Computer Science. Addison-Wesley (1979)

\bibitem{baudel04visCP}
Baudel, T., {\& al}:
\newblock {Visual CP, Reference Manual} (2004) Manufactured and distributed by
  Ilog, {\tt http://www2.ilog.com/preview/Discovery/}.

\bibitem{gjf04flairs}
Ghoniem, M., Jussien, N., Fekete, J.D.:
\newblock {VISEXP: Visualizing Constraint Solver Dynamics Using Explanations}.
\newblock In Barr, V., Markov, Z., eds.: Proc. of the 17th Int. of the Florida
  Artificial Intelligence Research Society Conference (FLAIRS'04), AAAI Press
  (2004)

\bibitem{DDG05}
Denmat, T., Ducass{\'e}, M., Ridoux, O.:
\newblock {Data Mining and Cross-Checking of Execution Traces. A
  reinterpretation of Jones, Harrold and Satsko test Information
  Visualization}.
\newblock RI 1743, IRISA, Rennes (France) (2005)

\bibitem{csmr2005}
Zaidman, A., Calders, T., Demeyer, S., Paredaens, J.:
\newblock {Applying Webmining Techniques to Execution Traces to Support the
  Program Comprehension Process}.
\newblock In Society, I.C., ed.: Proceedings of the 9th European Conference on
  Software Maintenance and Rengineering (CSMR 2005), Manchester, UK (2005)
  134--142

\end{thebibliography}

\end{document}